\documentclass[prb,aps,twocolumn,showpacs]{revtex4}
\usepackage{amsmath,bm}

\usepackage{graphicx}

\newcommand{\rl}{\rangle\!\langle}

\DeclareMathOperator{\im}{Im}
\begin{document}

\author{Anna Sitek}
\author{Pawe{\l} Machnikowski}
 \email{pawel.machnikowski@pwr.wroc.pl}
 \affiliation{Institute of Physics, Wroc{\l}aw University of
Technology, 50-370 Wroc{\l}aw, Poland}

\title{Interplay of coupling and superradiant emission in the optical
response of a double quantum dot}

\begin{abstract}
We study theoretically the optical response of a double quantum dot
structure to an 
ultrafast optical excitation. We show that the interplay of a
specific type of coupling between the dots and their collective
interaction with the radiative environment leads to very
characteristic features in the time-resolved luminescence as well as
in the absorption spectrum of the system. For a sufficiently strong
coupling, these effects survive even if the transition energy
mismatch between the two dots exceeds by far the emission line width.
\end{abstract}


\maketitle

\section{Introduction}

Systems composed of two quantum dots (QDs) have attracted much attention in
recent years. Many theoretical and experimental results have
demonstrated that the physical properties of such double quantum dots
(DQDs) are much richer than those of individual ones. 
On the one hand, this may pave the way to new applications, including
long-time storage of 
quantum information \cite{pazy01b}, conditional optical
control \cite{unold05} that may lead to an implementation of a two-qubit quantum
gate \cite{biolatti00}, generation of entangled photons
\cite{gywat02} or coherent optical spin control and entangling
\cite{troiani03,nazir04,gauger08b}. On the other hand, in order to
take advantage of these extended possibilities, the properties of DQDs
have to be understood and controlled. 

There are two major factors that determine the physical (in
particular, optical) properties of DQDs: the coupling between the dots
and their interaction with the environment. In both these areas new
features appear, as compared to the physics of individual dots. Both
problems have been in the focus of extensive experimental and
theoretical work but many important questions remain open.

Coupling between the dots is essential for quantum conditional
control, entanglement formation and implementation of two-qubit
gates. It is therefore understandable that considerable experimental
effort has been devoted to demonstrating its existence in double dot
systems \cite{bayer01,ortner03,ortner05,ortner05b,krenner05b}, while
theoretical studies were aimed at characterizing its signatures in the
optical response of DQDs \cite{danckwerts06}. 

Coupling of DQDs to their environment is affected by collective
effects, which may lead to superradiance phenomena
\cite{scheibner07}. Theoretical studies on the dephasing (in
particular, decay of entanglement) in double dot systems have shown
that coherence properties strongly depend on whether the
dots are coupled to a common reservoir or to separate reservoirs
\cite{yu02,yu03,tolkunov05,roszak06a}. The collective nature of the
coupling to the environment allows one to construct arrays (collective quantum
bits) which are more resistant to decoherence than individual systems
\cite{zanardi98b,grodecka06}. In fact, in any such array of two-level systems
collectively coupled to their common reservoir, the dephasing of some
states is slowed down, while other states decohere faster. In the case
of radiative decay, an
essential role is played by the transition energy mismatch between the
systems forming the array \cite{sitek07a}.  These  subradiance and
superradiance effects influence the optical response of DQDs and can
affect the coherence properties of DQD-based quantum devices.

Since the paper by Dicke \cite{dicke54}, coherent effects in the
radiative decay of two or more atoms have been studied in numerous works. The
emission from identical
\cite{stephen64,hutchinson64,lehmberg70a,milonni74,agarwal74} and
non-identical \cite{varfolomeev71,milonni75} two-level systems has been
studied and methods suitable for the description of arrays of various
shapes have been developed \cite{freedhoff04,rudolph04}. Systems
formed by QDs share many features 
with those made of natural atoms. In particular, QD samples may be modelled
as ensembles of two-level systems with parallel transition dipoles,
corresponding to the fundamental optical (interband) transition between the
ground (`empty dot') state and the confined exciton state
(electron-hole pair in the dot), although one
should not expect the transition energies to be exactly matched in
these artificial objects. Nonetheless, these
semiconductor structures are specific with some respects. 

The spacing
between QDs in intentionally manufactured DQD systems is typically
of the order of nanometers, that is 2--3 orders of magnitude smaller
than the wave length of the resonantly coupled radiation. This allows
one to assume the Dicke limit of the coupling and neglect the
retardation effects, which were one of the major concerns in the
general theory \cite{milonni74,milonni75}. On the other hand, small
distance between the two dots precludes individual addressing of each
system by the exciting field so that only certain initial
states may be prepared. This means, in particular, that the
two-excitation (biexciton, i.e., one electron-hole pair in each QD of
the DQD structure) states must be included in the description (except for
the weak excitation limit). 

\begin{figure}[tb]
\includegraphics[width=55 mm]{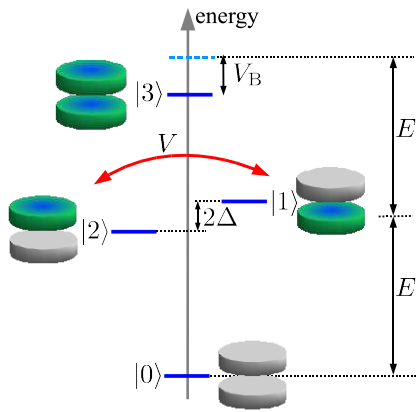}
\caption{\label{fig:diagram}(Color online) Graphical representation of
the energy levels (basis states) and couplings in the double quantum
dot system.} 
\end{figure}

Another important feature that distinguishes
QDs from atomic samples is the presence of two kinds of dipole
couplings. One of them is the direct
interaction between static dipole moments associated with the
electron-hole charge distribution in the two-dots \cite{unold05}. This kind of
coupling is obviously present only if the two dots are occupied by
excitons and its effect is to shift the energy
of the biexciton state (denoted by $V_{\mathrm{B}}$ in
Fig.~\ref{fig:diagram}). The other coupling is related to the
interband 
matrix elements of the electric dipole moment and is analogous to the
F\"orster coupling in molecular systems as well as to dipole couplings
between atoms which appear in the description of superradiance
phenomena in atomic samples. It couples the two single-exciton states
of the system via a process that may be imagined as a recombination of
the exciton in one of the dots with the subsequent transfer of energy
to the other one (via Coulomb interaction), where the exciton is
recreated (this coupling is denoted by $V$ in Fig.~\ref{fig:diagram}). 
Thus, this coupling
has an ``excitation transfer'' character. In QDs, this interaction
is modified by the finite size of the charge distribution (the QD size is
comparable with the inter-dot distance) and has a finite value in the
formal limit of vanishing distance \cite{govorov03,rozbicki08a,machnikowski09a}. 
Therefore, it is not a universal
function of the distance. Finally, apart from this dipole-dipole
coupling, other kinds of interaction may be present (e.g., effective
`tunneling' coupling accounting for a slight overlap of wave functions
which may dominate over the F\"orster coupling for closely spaced dots). For
these reasons, the strength of the coupling between the dots becomes an
essentially independent parameter. All in all, there are three
independent parameters governing the radiative properties of a double-dot
system in the Dicke limit, compared to two in the case of atoms.

In this paper we study the interplay of 
all the factors affecting the optical response of a DQD:
the coupling between the two dots
forming the DQD, the mismatch between their transition energies and
their collective interaction with the radiation 
environment (vacuum). With the recent experimental progress
\cite{vamivakas09,flagg09},  optical studies of
a single, resonantly driven nanostructure have become
feasible. Therefore, we study the simplest optical property of the
system, that is, the response to a single ultrafast pulse tuned to the
fundamental interband transition of the system. The
analysis includes the linear regime, where the spectral
properties of the optical response yield the absorption spectrum of
the system, as well as the higher order contributions which are
affected by the exciton-biexciton dynamics. We show that both the
time-resolved signal and its frequency spectrum can show clear
signatures of collective coupling to the radiation
reservoir. Moreover, in the realistic case of non-identical dots, the
striking features related to collective radiative relaxation appear
only when the inter-dot coupling is strong enough and has the
``excitation transfer'' character (as opposed to the
occupation-preserving biexciton shift). In this way, our results
provide a sensitive test for the appearance of a specific kind of
coupling between the dots.

The paper is organized as follows. In Sec. \ref{sec:system}, we
describe the system under study and define its
model. In Sec. \ref{sec:evol}, equations describing the evolution of
the optical polarization are derived. Sec. \ref{sec:discussion}
contains the discussion of the results. Summary and final remarks are
contained in Sec. \ref{sec:concl}. 

\section{The system}
\label{sec:system}

The system under study is a DQD composed of two QDs
placed at a distance much smaller than the relevant photon
wavelength. We restrict the discussion to the ground exciton states in
the two dots. Due to strong electron-hole Coulomb attraction, the
`spatially direct' states (electron-hole pairs confined in the same
dot) have much lower energy than the `dissociated' states (we do not consider
external electric fields which would change this picture
\cite{szafran05,szafran08}). Therefore, we include only the `spatially
direct' states in our model.
 We assume also that the polarization of the laser pulse
corresponds to a polarization eigenstate of the excitons, which allows
us to include only one out of the two bright states in each dot and
to describe the DQD as a four level system, with the state
$|0\rangle$ representing empty dots, $|1\rangle$ and $|2\rangle$
denoting the single exciton states with an exciton localized in the
first and in the second dot, respectively, and $|3\rangle$
representing the ``molecular biexciton'' state, that is, the state with both dots
occupied by an exciton. We denote the energies corresponding to the fundamental
optical transition in the two dots by $E_{1,2}=E\pm\Delta$ and allow
for a coupling between the dots, whose amplitude is $V$. The latter
may originate either from the interband dipole (F\"orster) coupling
between the dots \cite{danckwerts06} or appear as an effective description of tunnel
coupling if the carrier wave functions in the two dots overlap. In
addition, excitons confined in the two dots interact via their static dipole
moments, which shifts the energy of the biexciton state by
$V_{\mathrm{B}}$. The Hamiltonian describing the isolated DQD system is then
\begin{eqnarray*}
\tilde{H}_{\mathrm{X}} & = & 
(E+\Delta)|1\rl 1|+(E-\Delta)|2\rl 2|+(2E+V_{\mathrm{B}})|3\rl 3|\\
&&+V (|1\rl 2|+|2\rl 1|).
\end{eqnarray*}
The structure of energy levels is shown in Fig.~\ref{fig:diagram}.

The two dots interact with their common radiative reservoir. The
Hamiltonian describing this interaction is
\begin{equation*}
\tilde{H}_{\mathrm{int}}
=i\Sigma_{-}
\sum_{\bm{k},\lambda}\bm{d}\cdot\hat{e}_{\lambda}
\left(\bm{k}\right)\sqrt{\frac{\hbar\omega_{\bm{k}}}{2\epsilon_{0}\epsilon_{r}v}}
b^{\dagger}_{\bm{k},\lambda}+\mathrm{H.c.},
\end{equation*}
with 
$\Sigma_{-}=\Sigma_{+}^{\dag}=
(\sigma^{\left(1\right)}_{-}+\sigma^{\left(2\right)}_{-})$,
where
$\sigma^{\left(1\right)}_{-}=|0\rl 1|+|2\rl 3|$ and 
$\sigma^{\left(2\right)}_{-}=|0\rl 2|+|1\rl 3|$ are annihilation
operators for an exciton in the first and second dot, respectively, 
$\bm{k}$ is a photon wave vector, $\lambda$ denotes polarizations,
$b_{\bm{k},\lambda},b_{\bm{k},\lambda}^{\dag}$ are photon 
annihilation and creation operators,
$\bm{d}$ is the interband dipole moment (for simplicity equal for 
both QDs), $\hat{e}_{\lambda}\left(\bm{k}\right)$ is a unit polarization 
vector, $\epsilon_{0}$ is the vacuum permittivity,
$\epsilon_{\mathrm{r}}$ is the dielectric constant of the semiconductor,
and $v$ is the normalization volume for the EM modes. 

Finally,
\begin{equation*}
H_{\mathrm{phot}}=\sum_{\bm{k},\lambda}\hbar\omega_{{\bm{k}}}
b_{\bm{k},\lambda}^{\dag}b_{\bm{k},\lambda}
\end{equation*}
is the Hamiltonian of the photon reservoir, where
$\omega_{\bm{k}}$ is the frequency of the photon with a wave vector
$\bm{k}$. 

We will describe the evolution in a ``rotating basis'' defined by the
unitary transformation 
\begin{equation*}
U=
e^{iE(|1\rl 1|+|2\rl 2|+2|3\rl 3|)t/\hbar+iH_{\mathrm{phot}}t/\hbar}.
\end{equation*}
The transformed Hamiltonian is
\begin{equation*}
H = U(\tilde{H}_{\mathrm{X}}+\tilde{H}_{\mathrm{int}}
+H_{\mathrm{phot}})U^{\dag}+i\hbar \frac{dU}{dt}U^{\dag}
= H_{\mathrm{X}}+H_{\mathrm{int}},
\end{equation*}
where
\begin{eqnarray}\label{hamX}
H_{\mathrm{X}}=
\Delta(|1\rl 1|-|2\rl 2|)+V_{\mathrm{B}}|3\rl 3|
+V (|1\rl 2|+|2\rl 1|)
\end{eqnarray}
and
\begin{equation*}
H_{\mathrm{int}}
=i\Sigma_{-}
\sum_{\bm{k},\lambda}\bm{d}\cdot\hat{e}_{\lambda}
\left(\bm{k}\right)\sqrt{\frac{\hbar\omega_{\bm{k}}}{2\epsilon_{0}\epsilon_{r}v}}
e^{i\left(\omega_{\bm{k}}-E/\hbar\right)t}
b^{\dagger}_{\bm{k},\lambda}+\mathrm{H.c.}
\end{equation*}

\section{The system evolution and the optical response}
\label{sec:evol}

In this section we will describe the evolution of the DQD system after
an instantaneous excitation with an ultrashort pulse. An analysis of
both the time-resolved and spectrally-resolved response will be
performed. In the linear response limit, the latter provides the
linear susceptibility from which the absorption spectrum can be derived.

The source of the optical signal from the QD system is the electric
dipole moment (polarization) related to the interband
transition. Assuming identical 
magnitude and orientation of the transition dipoles in both dots, the
relevant (interband) part 
of the dipole moment operator can be written as  
\begin{equation*}
\hat{\bm{d}}=\bm{d}\Sigma_{-}+\mathrm{H.c.} 
\end{equation*}
The magnitude of the positive frequency part of the emitted coherent
optical field, normalized to its initial value is, therefore, 
\begin{equation}
P(t)=i\frac{\rho_{10}(t)+\rho_{20}(t)+\rho_{31}(t)+\rho_{32}(t)}{
\rho_{10}(0)+\rho_{20}(0)+\rho_{31}(0)+\rho_{32}(0)},
\label{polariz}
\end{equation}
where $\rho$ is the density matrix of the four-level DQD system and
$\rho_{ij}(t)=\langle i|\rho(t) |j\rangle$.
The matrix elements $\rho_{10}$ and $\rho_{20}$ are related to the
coherences between the ground state and the single exciton states, hence
are referred to as exciton polarizations. The other two matrix
elements, $\rho_{31}$ and $\rho_{32}$, correspond
to the transition between the single exciton and biexciton states and
are commonly called biexciton polarizations.

An ultrafast pulse is assumed to be spectrally broad enough in order
not to discriminate between the two dots in the structure. Due to a
small (sub-wavelength) distance between the dots, they cannot be
resolved spatially, either. Therefore, the pulse induces optical
polarizations symmetrically and independently in both dots. 
After the pulse, the
relevant elements of the density matrix have the values
\begin{subequations}
\begin{eqnarray}\label{initial}
\rho_{10}(0)=\rho_{20}(0) & = &
-\frac{i}{2}\sin\alpha\cos^{2}\frac{\alpha}{2}, \\
\rho_{31}(0)=\rho_{32}(0) & = &
-\frac{i}{2}\sin\alpha\sin^{2}\frac{\alpha}{2},
\end{eqnarray}
\end{subequations}
where $\alpha$ is the pulse area.

After this instantaneous initial excitation, the density matrix evolves
according to the Liouville--von Neumann--Lindblad equation of motion
\begin{equation}\label{lind}
\dot{\rho}=-\frac{i}{\hbar}\left[H_{\mathrm{X}},\rho\right]
+\mathcal{L}[\rho],
\end{equation}
with
\begin{equation}
\mathcal{L}[\rho]=\Gamma \left[ \Sigma_{-}\rho\Sigma_{+}
-\frac{1}{2}\{\Sigma_{+}\Sigma_{-},\rho \}_{+}\right],
\label{liouv}
\end{equation}
where 
\begin{equation*}
\Gamma=\frac{E^{3}|\bm{d}|^{2}\sqrt{\epsilon_{\mathrm{r}}}}{
3\pi\epsilon_{0}c^{3}\hbar^{4}}
\end{equation*}
is the spontaneous decay rate. 
It can be noted that, from the formal point of view, the evolution
equation (\ref{lind}) is not unique. In fact, a standard microscopic
derivation \cite{breuer02} involves grouping the transitions into sets
characterized by the values of the transition energies and, therefore,
produces two different Lindblad equations, depending on the arbitrary
classification of the two radiating systems as ``identical'' or
``different''. Therefore, it does not provide any means
for the study of the transition from the independent decay to
the collective regime as the transition energy mismatch $\Delta$ is
varied from zero to a finite value. On the other hand,
Eq.~(\ref{lind}), composed of the 
unitary and dissipative parts, is compatible with the Wigner-Weisskopf
description of the double QD system \cite{varfolomeev71,sitek07a},
where the Markov 
approximation is performed without any arbitrariness (assuming 
$\Delta\ll E$, so that the
reservoir spectral density does not vary considerably over the
relevant frequency range).

Note that the collective coupling of the two dots to the
electromagnetic field, as described by Eq.~(\ref{liouv}), results in
the appearance of cross-terms of the form
$\sigma_{-}^{(i)}\rho\sigma_{+}^{(j)}$, $i\neq j$, which are absent in
the case of individual coupling to separate reservoirs (see
Appendix). These terms correspond to the imaginary part of the
coupling that appears in other approaches
\cite{varfolomeev71,freedhoff04,rudolph04}. Thus, in the limit of vanishing
biexciton shift and (in the case of
Refs.~\onlinecite{freedhoff04,rudolph04}) of identical dots, our model is fully
equivalent to the small-system limit of those theories.  

The equation of motion (\ref{lind}) leads to the closed system of four
equations for the matrix elements of interest,
\begin{subequations}
\begin{eqnarray}\label{evol-eq}
\dot{\rho}_{10} &= &\left(-i\frac{\Delta}{\hbar}-\frac{\Gamma}{2} \right)\rho_{10}
+\left( -i\frac{V}{\hbar}- \frac{\Gamma}{2}\right) \rho_{20}\nonumber \\
&&+\Gamma(\rho_{31}+\rho_{32}), \\
\dot{\rho}_{20} &= &\left(-i\frac{V}{\hbar}- \frac{\Gamma}{2}\right)\rho_{10}
+ \left(i\frac{\Delta}{\hbar}-\frac{\Gamma}{2} \right)\rho_{20}\nonumber \\
&&+\Gamma(\rho_{31}+\rho_{32}), \\
\dot{\rho}_{32} &=
&\left(-i\frac{V_{\mathrm{B}}}{\hbar}-i\frac{\Delta}{\hbar}
-\frac{3\Gamma}{2} \right)\rho_{32} \nonumber \\
&&+\left(i\frac{V}{\hbar}- \frac{\Gamma}{2}\right) \rho_{31},\\
\dot{\rho}_{31} &= &\left(i\frac{V}{\hbar}- \frac{\Gamma}{2}\right)\rho_{32}\nonumber\\
&&+
\left(-i\frac{V_{\mathrm{B}}}{\hbar}+i\frac{\Delta}{\hbar}-\frac{3\Gamma}{2} 
\right)\rho_{31}.
\end{eqnarray}
\end{subequations}
This system of equations with the initial values given by
Eqs.~(\ref{initial}-b) can be solved by the standard Laplace transform
method. Then, for the total emitted signal [Eq. (\ref{polariz})], we find
\begin{equation}\label{pol-norm}
P(t)=\frac{i}{2}\sum_{i}\left[ 
\cos^{2}\frac{\alpha}{2} A_{i}+ \sin^{2}\frac{\alpha}{2}B_{i}\right]
e^{\lambda_{i}t},
\end{equation}
where (with the first index corresponding to the upper sign)
\begin{subequations}
\begin{eqnarray}
\lambda_{1,2} & = & \pm i\tilde{\Omega}_{-}-\frac{\Gamma}{2}, \label{lambda12}\\
\lambda_{3,4} & = & i(\pm\tilde{\Omega}_{+}-V_{\mathrm{B}}/\hbar)
	-\frac{3\Gamma}{2}, \\
A_{1,2} & = & 1\mp\frac{V/\hbar-i\Gamma/2}{\tilde{\Omega}_{-}},\label{A12}\\
B_{1,2} & = & 
\frac{\mp 2\Gamma V_{\mathrm{B}}\tilde{\Omega}_{-}+2\Gamma V_{\mathrm{B}}V/\hbar
-i\Gamma^{2}V_{\mathrm{B}}-2\Delta^{2}\Gamma/\hbar}{
\tilde{\Omega}_{-}[\mp 2\Gamma V-(iV_{\mathrm{B}}+\hbar\Gamma)
(2\tilde{\Omega}_{-}\pm V_{\mathrm{B}}/\hbar\mp i\Gamma)]},\label{B12}\\
A_{3,4} & = & 0, \\
B_{3,4} & = &
\frac{\pm (V_{\mathrm{B}}/\hbar-2i\Gamma)(2iV-iV_{\mathrm{B}}+\hbar\Gamma)
+2iV_{\mathrm{B}}\tilde{\Omega}_{-}}{
(iV_{\mathrm{B}}+\hbar\Gamma)(2\tilde{\Omega}_{+}
\mp V_{\mathrm{B}}/\hbar \pm i\Gamma )\pm 2\Gamma V} \nonumber\\
&& +\frac{
4\Gamma V^{2}/\hbar-iV_{\mathrm{B}}^{2}V/\hbar^{2}
+\Gamma V_{\mathrm{B}}^{2}/(2\hbar)+\hbar\Gamma^{3}}{
\tilde{\Omega}_{+}[(iV_{\mathrm{B}}+\hbar\Gamma)
(2\tilde{\Omega}_{+}\mp V_{\mathrm{B}}/\hbar \pm i\Gamma )
\pm 2\Gamma V]},
\end{eqnarray}
\end{subequations}
with 
\begin{equation}\label{Omega-pm}
\tilde{\Omega}_{\pm}=\sqrt{(V/\hbar)^{2}\pm iV\Gamma/\hbar
-\Gamma^{2}/4+(\Delta/\hbar)^{2}}.
\end{equation}
The Fourier transform of this signal is
\begin{equation}\label{P-w}
\hat{P}(\omega)=\frac{\cos^{2}\frac{\alpha}{2}f(\omega)
+\sin^{2}\frac{\alpha}{2}g(\omega)}{(\omega-i\lambda_{1})(\omega-i\lambda_{2})
(\omega-i\lambda_{3})(\omega-i\lambda_{4})},
\end{equation}
where 
\begin{equation}\label{f-w}
f(\omega) =  i(\omega+V/\hbar)(\omega-i\lambda_{3})(\omega-i\lambda_{4})
\end{equation}
and
\begin{eqnarray*}
g(\omega) & = & i(\omega-V_{\mathrm{B}}/\hbar-V/\hbar+3i\Gamma)
(\omega-i\lambda_{1})(\omega-i\lambda_{2})\\
&&+2\Gamma V_{\mathrm{B}}(V/\hbar+\omega)/\hbar-2\Delta^{2}\Gamma/\hbar^{2}.
\end{eqnarray*}
It should be noted that the frequency $\omega$ in the above equations
(and in the following discussion) is defined with respect to the mean
transition frequency, that is, $\omega=\tilde{\omega}-E/\hbar$, where
$\tilde{\omega}$ is the actual frequency of the emitted
electromagnetic radiation. 

\section{Discussion}
\label{sec:discussion}

The evolution of the optical polarization after an instantaneous
excitation depends on whether the dots are coupled or not. In this
section we will discuss the two cases, comparing the optical response of the
two dots interacting with the common reservoir in the Dicke limit
(separation of the dots is small compared to $\hbar c/E$),
using the solution derived in Sec.~\ref{sec:evol}, with the response
of a hypothetical system consisting of two dots interacting with
independent reservoirs. 

\subsection{Uncoupled dots}
\label{sec:uncoupled}

\begin{figure}[tb]
\includegraphics[width=85 mm]{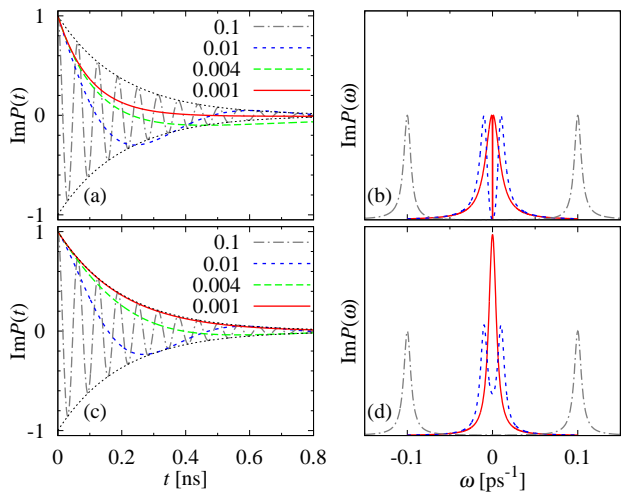}
\caption{\label{fig:v0b0}(Color online) (a,c) The optical polarization as a function
of time after an ultrafast excitation for uncoupled
dots ($V=V_{\mathrm{B}}=0$) in 
the linear response limit ($\alpha\to 0$). (b,d) The corresponding
spectrum. Figs. (a,b) show the results for QDs interacting with a
common reservoir in the Dicke limit, while Figs. (c,d) refer to dots
radiating into independent reservoirs. Here $\Gamma=0.01$~ps$^{-1}$
and the values of $\Delta/\hbar$ are as shown in the figure (in ps$^{-1}$). Line
definitions in (b,d) are the same as in (a,c). Dotted lines
in (a,c) show the envelope $\pm\exp(-\Gamma t/2)$. The vertical scale
in (b) and (d) is the same.}
\end{figure}

In the case of uncoupled QDs ($V=V_{\mathrm{B}}=0$) the optical response is
determined by the interplay of the other two parameters: the
recombination rate $\Gamma$ and the transition energy mismatch
$\Delta$. In Fig.~\ref{fig:v0b0} we show the optical polarization in
the time and frequency domain for a fixed value of the radiative
recombination time  $1/\Gamma=100$~ps. We assume here $\alpha\to 0$
(linear response limit), so that the imaginary part of the Fourier
transform of the normalized signal is proportional to the absorption
spectrum of the DQD. In this case, $P(t)$ is purely imaginary.

For $\Delta\gg \hbar\Gamma$ the evolution of the
optical signal is dominated by optical beats due to the interference
of fields emitted from the two dots [Fig.~\ref{fig:v0b0}(a,c), gray
dash-dotted line]. There is no noticeable difference
between the cases of a common reservoir [Fig.~\ref{fig:v0b0}(a)] and
separate reservoirs [Fig.~\ref{fig:v0b0}(c)]. This is not surprising
since systems with different transition energies emit into different
frequency sectors of the reservoir and thus essentially interact with
different reservoirs anyway. The two spectra
[Fig.~\ref{fig:v0b0}(b,d)] also look indistinguishable in this case.

This situation changes as the
energy mismatch $\Delta$ is decreased (Fig.~\ref{fig:v0b0}, blue
and green dashed lines). 
One effect, which follows from
Eqs.~(\ref{lambda12}) and (\ref{Omega-pm}), is that the frequency is
decreased in the collective case, although this does not lead to
any qualitative difference in the evolution of the optical polarization
represented in the time domain [Fig.~\ref{fig:v0b0}(a,c)]. A much more
pronounced difference is visible in the spectra
[Fig.~\ref{fig:v0b0}(b,d)]. From Eqs.~(\ref{P-w}) and (\ref{f-w}) one
has in the present special case, for $\Delta>\hbar\Gamma/2$
\begin{eqnarray*}
\lefteqn{\im P(\omega|\alpha=0,V=0,\Delta>\hbar\Gamma/2)=}\\
&&\frac{\Gamma\omega/\Omega_{+}}{4(\omega-\Omega_{+})^{2}+\Gamma^{2}/4}
-\frac{\Gamma\omega/\Omega_{+}}{4(\omega+\Omega_{+})^{2}+\Gamma^{2}/4},
\end{eqnarray*}
where $\Omega_{+}=\sqrt{(\Delta/\hbar)^{2}-\Gamma^{2}/4}$. This takes the form
of two Lorentzians centered around $\omega=\pm\Omega_{+}$ 
only as long as $\Gamma/\Omega_{+}$ is small
and only for $|\omega-\Omega_{+}|\lesssim\Gamma$. 
In fact, $P(\omega=0)=0$, which means that the spectrum
must considerably differ from the sum of two Lorentzians when the
latter overlap. This can be seen in Figs.~\ref{fig:v0b0}(b,d) (blue
dashed line). 

The particular features of the DQD spectrum in the common reservoir
case become even more pronounced when $\Delta<\hbar\Gamma/2$. Now
$\Omega_{+}$ is imaginary and the absorption spectrum can be written
as 
\begin{eqnarray*}
\lefteqn{\im P(\omega|\alpha=0,V=0,\Delta<\hbar\Gamma/2)=}\\
&&\frac{\Gamma_{+}}{|\Omega_{+}|}\frac{\Gamma_{+}}{4\omega^{2}+\Gamma_{+}^{2}}
-\frac{\Gamma_{-}}{|\Omega_{+}|}\frac{\Gamma_{-}}{4\omega^{2}+\Gamma_{-}^{2}},
\end{eqnarray*}
where $\Gamma_{\pm}=\Gamma\pm 2|\Omega_{+}|$.
Since $\Gamma_{+}\to 2\Gamma$ as $\Delta\to 0$, the  width of the first
Lorentzian becomes twice larger than it was for independent
reservoirs. This term corresponds to the superradiant component of the
system evolution \cite{sitek07a}. The second Lorentzian (which is
negative) becomes narrow, since $\Gamma_{-}\to 0$ as $\Delta\to 0$,
and represents the subradiant component. However, its weight
vanishes in the limit of identical dots [note that the amplitudes of
the two Lorentzians are always exactly opposite, in accordance with
$P(\omega=0)=0$]. These spectral features are reflected by the
evolution in the time domain shown in Fig.~\ref{fig:v0b0}(a). For
$\Delta$ slightly below $\hbar\Gamma/2$ (green long dashed line), the
evolution is a sum of two exponential factors: one positive, large,
and short living and the other one negative, small, and long
living. In fact, however, the resulting behavior is hardly
distinguishable from the strongly damped oscillations appearing for
individually emitting dots [Fig.~\ref{fig:v0b0}(c)]. Only for
$\Delta\ll \hbar\Gamma$ the difference becomes remarkable: in the case of
collective emission, the polarization becomes dominated by the
superradiant component which leads to a decay with a doubled rate, as
compared to the separate reservoir case [red
solid lines in Figs.~\ref{fig:v0b0}(a,c)].

\begin{figure}[tb]
\includegraphics[width=85 mm]{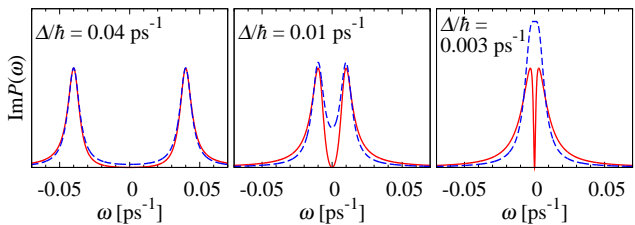}
\caption{\label{fig:transit}(Color online) The transition from the
  `different dots' regime to the `identical dots' regime as manifested
  by the form of the absorption spectrum of a DQD system.  Here $\Gamma=0.01$~ps$^{-1}$
and the values of $\Delta/\hbar$ are shown in the figures. Red solid
lines: the actual system; blue dashed lines: hypothetical system made of
two dots interacting with separate reservoirs.}
\end{figure}

These results allow one to understand the transition from the limit of
different systems (separate reservoirs) to identical systems (common
reservoir). This transition is manifested in the reconstruction of the
absorption spectrum, as shown in Fig.~\ref{fig:transit}. As long as $\Delta\gg
\hbar\Gamma/2$, the two dots are
coupled to different frequency sectors of the electromagnetic
reservoir and the spectrum of the DQD system is almost
indistinguishable from that of two dots coupled to separate
reservoirs. In the most interesting parameter range,
$\Delta\sim\hbar\Gamma/2$, the spectrum is 
non-Lorentzian as long as
$\Delta>\hbar\Gamma/2$ and then switches to an unusual form of two
Lorentzians with different weights and opposite signs, centered at
zero frequency. Only then the evolution of polarization has two
exponentially decaying components. 

Although the features discussed above are interesting from a general
physical point of view, their appearance requires that the energy
mismatch between the dots is comparable with the radiative line width,
that is, of the order of tens of $\mu$eV. In spite of the rapid
progress of nanostructure manufacturing, this can be very hard to
achieve experimentally, unless the dots can be driven to resonance by
external fields (e.g. taking advantage of a different strength of the
DC Stark effect in the two dots). As we will show in
Sec. \ref{sec:coupled-V}, 
if the dots are coupled by an ``excitation transfer''
interaction, the collective effects manifest themselves already for
much larger values of the energy mismatch.
Before we proceed to this case, we study the effect of the
other type of long-range interaction between confined excitons which
is due to static electric dipoles and results in a biexciton shift.

\subsection{Coupled dots: Biexciton shift}
\label{sec:coupled-biex}

\begin{figure}[tb]
\includegraphics[width=85 mm]{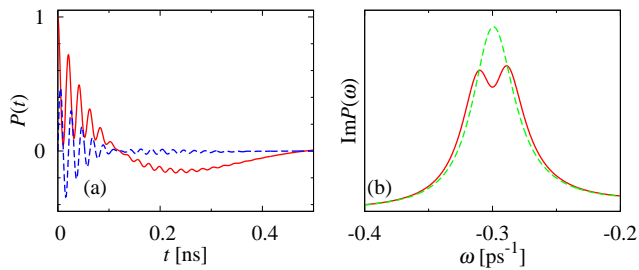}
\caption{\label{fig:v0b05}(Color online) (a) The optical polarization as a function
of time after an ultrafast excitation for dots coupled by the static
dipole moments only ($V=0$, $V_{\mathrm{B}}\neq 0$) for $\tan^{2}(\alpha/2)=0.2$,
$\Gamma=0.01$~ps$^{-1}$, $\Delta/\hbar=0.01$~ps$^{-1}$, 
$V_{\mathrm{B}}/\hbar=-0.3$~ps$^{-1}$.
Blue dashed line: real part,
red solid line: imaginary part. (b) The Fourier transform of the
polarization signal. Red solid line: parameters as in (a). Green
dashed line: $\Delta/\hbar=0.001$~ps$^{-1}$.}
\end{figure}

In this section we discuss the evolution of the optical polarization
for a system of two QDs coupled by a static dipole interaction 
$V_{\mathrm{B}}$ (we keep $V=0$). 
This coupling is important only for
the third (and higher) order terms in the optical signal. 
In the parameter range where the
transition between the two regimes of evolution occurs, as discussed
in Sec.~\ref{sec:uncoupled}, the biexciton shift $V_{\mathrm{B}}$
(which is of the order of meV) is likely to be the largest energy
scale of the problem. Therefore, we restrict the discussion to the case
of $V_{\mathrm{B}}\gg\Delta$ and $V_{\mathrm{B}}\gg\hbar\Gamma$. 
Then, from Eq.~(\ref{B12}), in the leading order,
$B_{1,2}\sim\hbar\Gamma/V_{\mathrm{B}}$, which means that
the correction to the components evolving with the frequencies
$i\lambda_{1}$ and $i\lambda_{2}$, as well as to the spectrum of the
optical response in the frequency range studied in
Sec.~\ref{sec:uncoupled}, is negligible.  

In the spectral region of $\omega\sim V_{\mathrm{B}}/\hbar$, a new feature
appears when $V_{\mathrm{B}}\neq 0$ [see Fig.~\ref{fig:v0b05}(b)]. 
The shape of this feature
evolves in a different way, as compared to the absorption line at $\omega=0$. For
$\Delta\gg\hbar\Gamma$ two nearly Lorentzian lines of almost equal magnitude
appear at $\omega=(V_{\mathrm{B}}\pm\Delta)/\hbar$. These two lines collapse
as $\Delta\sim\hbar\Gamma$ but in this case they retain their approximately
Lorentzian shape. For $\Delta<\hbar\Gamma/2$
both Lorentzians are centered at the same frequency
$\omega=V_{\mathrm{B}}/\hbar$, as in the absorption spectrum discussed
above, but now their
widths tend to $\Gamma$ and $2\Gamma$ as $\Delta\to 0$. The area of
the narrower line (which is negative) vanishes in this limit which, in
view of the finite width, implies that the amplitude of this line
vanishes. On the other hand, the area of the broader line remains
finite. This means that the evolution of the biexciton beats [presented in
Fig.~\ref{fig:v0b05}(a)] shows a two-rate damping which becomes
dominated by the short-living component as the dots become
identical. In contrast, as follows from the formulas listed in the
Appendix, in the case of independent reservoir the polarization decay rate is
always $3\Gamma/2$.

\subsection{Coupled dots: transfer interaction}
\label{sec:coupled-V}

In this section we consider the case of two dots coupled by an ``excitation
transfer'' interaction, that is, $V\neq 0$ in Eq.~(\ref{hamX}). We
assume that the magnitude of this coupling is larger than the
relaxation rate, $V\gg \hbar\Gamma$. As we will see, values of the
energy 
mismatch $\Delta$ for which the transition to the
collective behavior takes place is now of the order of $V$ so that
$\Gamma$ becomes the smallest frequency in the problem. 
One can, therefore, simplify the discussion by retaining only the
terms up to linear order in $\Gamma$ in Eqs.~(\ref{lambda12},b) and 
taking the amplitudes given by Eqs.~(\ref{A12}-f) at $\Gamma=0$. 

Then one gets
$\lambda_{1,2}=\pm i\Omega-\Gamma_{\mp}/2$ and 
$A_{1,2}=1\mp V/(\hbar\Omega)$,
where
$\Gamma_{\mp}=[ 1\mp V/(\hbar\Omega)] \Gamma$,
$\Omega=[\Delta^{2}+V^{2}]^{1/2}/\hbar$.
Thus, in the linear response limit, the system shows a decay described by two
exponential components. The subradiant one decays with the rate 
$\Gamma_{-}/2<\Gamma/2$ and its amplitude vanishes when $\Delta\ll V$. The
decay rate of the superradiant component is $\Gamma_{+}/2>\Gamma/2$. This
component dominates the decay in the limit of strongly coupled dots. 

\begin{figure}[tb]
\includegraphics[width=85 mm]{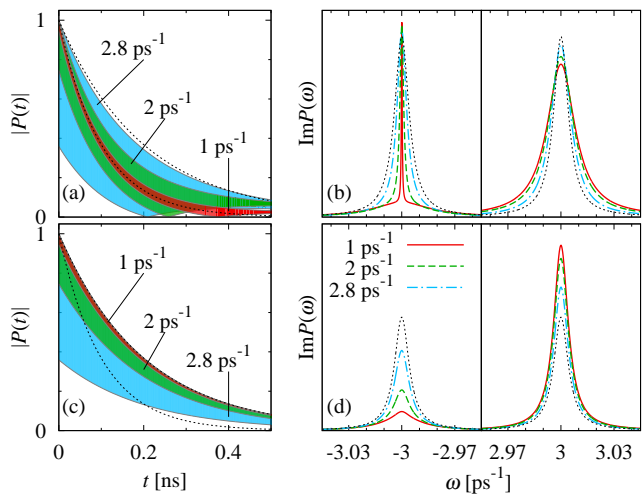}
\caption{\label{fig:w3b0}(Color online) Left panels: envelope of the optical beats
for two coupled QDs emitting to a common
reservoir and to separate reservoirs. For each value of $\Delta/\hbar$
(shown in the plot), the
coupling $V$ is adjusted so that $\Omega=3$~ps$^{-1}$
in each case. Right panels: the
corresponding absorption spectra. Plots (a) and (b) show the
properties of a DQD emitting to a common reservoir, while plots (c,d)
correspond to two dots emitting to separate reservoirs. Dotted lines
in (a,c) show exponential decay curves with the rates $\Gamma/2$ and
$\Gamma$, while in (b,d) they show Lorentzians of equal weight and
width $\Gamma/2$.}
\end{figure}

The evolution of the optical polarization in the present case is shown
in Fig.~\ref{fig:w3b0}(a), where we plot the envelopes of the optical
beats for three different sets of parameters $\Delta$ and $V$. 
Two effects that may be noticed are the decrease of the beating
amplitude and the change in the decay rate of the signal as the energy
mismatch decreases and the coupling increases. 
By comparison with the case of independent reservoirs
[Fig.~\ref{fig:w3b0}(c)] one can see that the first of these two effects
appears in both cases. A non-selective excitation
induces a symmetric superposition of the two occupations which has a
larger overlap with one of the eigenstates of the coupled
dots. Therefore, one of the oscillators contributing to the beats has a
larger amplitude and the beating amplitude is reduced. 

In contrast, the other effect is related to the collective interaction
with the photon reservoir. In the physical case of a common reservoir,
the eigenstate dominating the response is the superradiant one. In the
limit of $\Delta\to 0$ only the superradiant state is excited (it
coincides with the optically active symmetric superposition). This is
reflected by the decay of the optical response, as shown in
Fig.~\ref{fig:w3b0}(a). For
$\Delta\gg V$, the optical polarization decays with the rate
$\Gamma/2$, characteristic of a single system. As the coupling $V$
becomes comparable with the energy mismatch $\Delta$, the decay
becomes non-exponential and, in fact, contains two components decaying
with different rates. When the coupling dominates over the energy
mismatch, $V\gg\Delta$, the subradiant component vanishes and the
signal decays with the doubled rate $\Gamma$. On the contrary,
hypothetical dots coupled to independent reservoirs
[Fig.~\ref{fig:w3b0}(c)] always show a decay of the optical response
with the same rate $\Gamma/2$.

The absorption spectrum corresponding to the time-resolved response
discussed above is shown in Figs.~\ref{fig:w3b0}(b,d). The two models of
common and separate reservoirs differ essentially. In both cases
there is a similar transfer of line weight from one line to the other
as the energy mismatch decreases and the coupling increases. 
However, the way the line shapes change is
very different. In the case of independent reservoirs
[Fig.~\ref{fig:w3b0}(d)], the line widths 
remain constant (no sub- and superradiance effects) and only the line
amplitudes change. On the contrary, when the dots are coupled to a
common reservoir, the amplitudes of the lines are almost constant but
their widths change. This is quite a remarkable signature of the joint
appearance of inter-dot coupling and collective radiative decoherence
in the system. It is also interesting to note that, since the weights
of each line behaves almost in the same way in both cases, no difference
can be observed if the absorption is averaged over an inhomogeneous
ensemble. 

\section{Conclusions}
\label{sec:concl}

We have shown that the coexistence of coupling between quantum dots
and collective effects in their interaction with the electromagnetic
environment leaves clear traces in the optical response of these
systems. 

For uncoupled dots, the collective radiative properties (sub-
and superradiance) become important only when the energy mismatch
falls below the absorption line width. The decay of the
polarization then evolves from damped oscillations (beats) for
different dots to a superradiant exponential decay for identical
dots. For very similar dots the corresponding absorption line is
composed of two Lorentzians superposed at the same transition
frequency, with the same amplitude but different 
widths and opposite signs. 

Coupling between the dots changes this picture considerably. Now
superradiance effects can be observed as long as the energy mismatch
is smaller or comparable to the coupling strength. The envelope of the
optical beats in this case decays with the usual rate for different
dots and with the double (superradiant) rate when the coupling becomes
much larger than the energy mismatch. In the intermediate range, the
decay is composed of two exponential components. 

These effects in the
time-resolved optical response are reflected in the absorption
spectrum, where the lines corresponding to the small and large
components in the optical response loose or gain, respectively, their
widths (reflecting the sub- and superradiance property). This presents
a clear difference with respect to the (hypothetical) case of dots emitting to
different reservoirs where the lines loose or gain amplitude, while
their widths remain constant. This essentially different form of
absorption lines which, for sufficiently strongly coupled dots, can be
observed even for a transition energy mismatch in the range of
milli-electron-volts provides an experimentally accessible
signature of coupling and collective decoherence in double quantum
dots.

\begin{acknowledgments}
This work was supported by the Polish MNiSW under Grant No. N N202 1336
33. 
\end{acknowledgments}

\appendix
\section{Radiative decay to independent reservoirs}

In this appendix, we derive the equations of motion for the optical
polarizations and find the resulting optical response in the case of a
two coupled QDs interacting with separate photon reservoirs.

The initial values for the relevant matrix elements (at $t=0$) are given by
Eqs.~(\ref{initial},b). At $t>0$, the density matrix evolves
according to Eq.~(\ref{lind}) but the Lindblad operator now has the form
\begin{eqnarray*}
	\mathcal{L}[\rho]=\sum_{i=1,2}\Gamma \left[ 
\sigma^{(i)}_{-}\rho\sigma^{(i)}_{+}
-\frac{1}{2}\{\sigma^{(i)}_{+}\sigma^{(i)}_{-},\rho \}_{+}\right].
\end{eqnarray*}
This equation corresponds to the system of evolution equations for the
optical coherences
\begin{eqnarray*}
\dot{\rho}_{10} &= &\left(-i\Delta/\hbar-\Gamma/2 \right)\rho_{10}
-i(V/\hbar) \rho_{20} +\Gamma\rho_{32}, \\
\dot{\rho}_{20} &= &-i(V/\hbar)\rho_{10}
+ \left(i\Delta/\hbar-\Gamma/2 \right)\rho_{20} +\Gamma\rho_{31}, \\
\dot{\rho}_{32} &= &\left(-iV_{\mathrm{B}}/\hbar- i\Delta/\hbar
-3\Gamma/2 \right)\rho_{32} +i(V/\hbar) \rho_{31},\\
\dot{\rho}_{31} &= &i(V/\hbar)\rho_{32}
+ \left(-iV_{\mathrm{B}}/\hbar+i\Delta/\hbar-3\Gamma/2 \right)\rho_{31}.
\end{eqnarray*}
By solving this system of equations one finds the normalized optical
polarization defined by Eq. (\ref{pol-norm}) with 
\begin{eqnarray*}
&&\lambda_{1,2}  =  \pm i\Omega-\frac{\Gamma}{2}, \quad
\lambda_{3,4}  =  i(\pm \Omega-V_{\mathrm{B}}/\hbar)
	-\frac{3\Gamma}{2}, \\
&&A_{1,2} =  1\mp\frac{V}{\hbar\Omega},\quad A_{3,4}  =  0, \\
&&B_{1,2}  =  
\frac{\Gamma [(V_{\mathrm{B}}/\hbar-i\Gamma)(\mp \Omega+V/\hbar)
-2\Delta^{2}/\hbar^{2}]}{
2\Omega[(-iV_{\mathrm{B}}/\hbar-\Gamma)
(\pm V_{\mathrm{B}}/\hbar\mp i\Gamma +2\Omega)]},\\
&&B_{3,4}  = \frac{(iV_{\mathrm{B}}+\hbar\Gamma)(V_{\mathrm{B}}-2V)
(V/\hbar\pm\Omega)+2\Delta^{2}V_{\mathrm{B}}/\hbar}{
\Omega(iV_{\mathrm{B}}+\hbar\Gamma)
(\pm V_{\mathrm{B}}\mp i\hbar\Gamma-2\hbar\Omega)}, \nonumber\\
\end{eqnarray*}
where $\Omega=\sqrt{V^{2}+\Delta^{2}}/\hbar$.
\begin{equation*}
\end{equation*}
The Fourier transform of this signal is
\begin{equation*}
P(\omega)=\frac{\cos^{2}\frac{\alpha}{2}f(\omega)
+\sin^{2}\frac{\alpha}{2}g(\omega)}{(\omega-i\lambda_{1})(\omega-i\lambda_{2})
(\omega-i\lambda_{3})(\omega-i\lambda_{4})},
\end{equation*}
where 
$f(\omega) =  i(\omega+V/\hbar+i\Gamma/2)(\omega-i\lambda_{3})(\omega-i\lambda_{4})$
and
\begin{eqnarray*}
g(\omega) & = &
i(\omega-V/\hbar-V_{\mathrm{B}}/\hbar+5i\Gamma/2)
(\omega-i\lambda_{1})(\omega-i\lambda_{2}) \\
&&+\Gamma(\omega+V/\hbar+i\Gamma/2)(V_{\mathrm{B}}/\hbar-i\Gamma)
-2\Delta^{2}\Gamma/\hbar^{2}.
\end{eqnarray*}


\end{document}